\documentclass[prb,twocolumn,longbibliography,nofootinbib]{revtex4-1}
\usepackage{graphicx}
\usepackage{bm}
\usepackage{wrapfig}
\usepackage{color}
\usepackage{natbib}
\usepackage{hyperref}
\usepackage{amssymb}
\usepackage{amsmath}
\usepackage{geometry}
\usepackage{comment}

\def\Xint#1{\mathchoice
   {\XXint\displaystyle\textstyle{#1}}%
   {\XXint\textstyle\scriptstyle{#1}}%
   {\XXint\scriptstyle\scriptscriptstyle{#1}}%
   {\XXint\scriptscriptstyle\scriptscriptstyle{#1}}%
   \!\int}
\def\XXint#1#2#3{{\setbox0=\hbox{$#1{#2#3}{\int}$}
     \vcenter{\hbox{$#2#3$}}\kern-.5\wd0}}

\def\dashint{\Xint-}

\begin{document}

\title{Current distribution in a slit connecting two graphene half-planes}

\author{Sergey S. Pershoguba$^{1,2}$, Andrea F. Young$^3$, and Leonid I. Glazman$^1$; }

\affiliation{$^1$Department of Physics, Yale University, New Haven, CT 06520, USA}

\affiliation{$^2$Department of Physics and Astronomy, University of New Hampshire, Durham, New Hampshire 03824, USA}

\affiliation{$^3$Department of Physics, University of California, Santa Barbara, CA 93106}

\date{\today}

\begin{abstract}
We investigate the joint effect of viscous and Ohmic dissipation on electric current flow through a slit in a barrier dividing  a graphene sheet in two. In the case of the no-slip boundary condition, we find that the competition between the viscous and Ohmic types of the charge flow results in the evolution of the current density profile from a concave to convex shape. We provide a detailed analysis of the evolution and identify favorable conditions to observe it in experiment. In contrast, in the case of the no-stress boundary condition, there is no qualitative difference between the current  profiles in the Ohmic and viscous limits. The dichotomy between the behavior corresponding to distinct boundary conditions could be tested experimentally. 
\end{abstract}

\maketitle

\section{Introduction}
Recent years have seen a revival of interest in the idea\cite{Gurzhi_1968} that charge transport in solids under some conditions is best described by hydrodynamic flow of an electron liquid. Graphene  provides an ideal platform for observing hydrodynamic effects due to the extremely long electron mean free path for impurity scattering~\cite{TorreGeim2015,Bandurin2016,Crossno2016,FalkovichLevitovNatPhys2016,Lukas2016,FalkovichLevitovPRL2017,Guo3068,guo2018a,Berdyugin162,Gallagher158,Lukas2018}. In constrained geometries, viscous electron flow differs  from both the Ohmic and ballistic transport regimes. The simplest manifestations of that difference are seen in the conductance: it exceeds the ballistic limit for a slit connecting two conducting half-planes~\cite{Guo3068}, and may become negative for certain configurations of contacts along the edge of a conducting stripe~\cite{Bandurin2016}. 
These manifestations are fairly insensitive to the type of boundary condition for the electron liquid flowing around obstacles. For example, the conductance of a slit in the hydrodynamic regime exceeds the ballistic limit, regardless the liquid ``sticking'' to the boundary or ``sliding'' along it.

Sticking to or sliding along the boundary corresponds, respectively, to the no-slip or no-stress boundary conditions for the electron liquid.
There is no consensus in the literature (see Ref.  [\onlinecite{TorreGeim2015}] vs [\onlinecite{Guo3068}]) regarding which boundary condition is appropriate for graphene. Theoretical work~[\onlinecite{KiselevSchmalian2019}] discussed the relation of the hydrodynamic boundary conditions to the microscopic\cite{Fuchs1938} conditions for electron scattering off the boundary. 

Recently, spatially resolved experimental techniques have made it possible to investigate the velocity distribution in the electron flow \cite{Young2020,ShahalIlani2019,Walsworth2019}, giving direct information about the boundary conditions for hydrodynamic charge carriers. That motivates us to investigate theoretically the effect of boundary conditions and of the Ohmic losses in the bulk on the on the velocity distribution. We focus on the electron flow through a slit, see Fig.~\ref{fig:3D_flow}(a).

Our main finding is that the velocity profile may allow one to unambiguously determine the type of boundary conditions as well as to identify the viscous regime. We also elucidate the domain for the sample parameters (the slit width, charge carrier density, and temperature) favoring the hydrodynamic regime.

We start with a brief review in Sec.~\ref{sec:hydrodynamics_review} of the continuous-medium (hydrodynamic) equations which account for the electron viscosity and Ohmic losses. In the same Section, we identify the width of the boundary layer defined by the competition between the viscous and Ohmic terms in the hydrodynamic equations. The comparison of the limiting cases where either the viscous or Ohmic term dominates allows us to conclude in Sec.~\ref{sec:limiting_cases} that in the case of no-stress boundary condition it may be hard to distinguish in an experiment between the Ohmic and viscous electronic flows. In contrast, for the no-slip boundary condition, we notice a qualitative feature: the current density profile  is concave and convex in the Ohmic and viscous limits, respectively. In practice, the viscous term in the dynamic equation for the electron liquid coexists with the Ohmic term. In Sec.~\ref{sec:crossover}, we study the crossover between the two regimes controlled by a single dimensionless parameter, the ratio of the slit width to the width of the boundary layer introduced in Sec.~~\ref{sec:hydrodynamics_review}. We find the current density profile numerically at any value of this control parameter and present a simplified model allowing for an analytical solution, which agrees well with the numerical results. The control parameter may be varied {\it in} situ by changing the electron density and temperature. We identify the domain of parameters favoring the hydrodynamic regime of electron flow and map out the crossover lines separating the Ohmic, viscous and ballistic regimes from each other in Sec.~\ref{Sec:parameterdomain}. We conclude in Sec.~\ref{sec:conclusion}.




\section{Hydrodynamic description of electronic flow in graphene.} \label{sec:hydrodynamics_review}

In this section, we set up hydrodynamic equations and briefly discuss their applicability. Following previous literature 
\cite{FalkovichLevitovNatPhys2016,TorreGeim2015,FalkovichLevitovPRL2017,Guo3068,Berdyugin162,Gallagher158},
the electronic flow in graphene may be described, at low applied bias, by the linearized Stokes equation in two dimensions $\bm r = (x,y)$:
\begin{align}
    [\eta\, \Delta - (ne)^2\rho\,]\,\bm v(\bm r) = ne \bm \nabla\phi(\bm r). \label{Stokes}
\end{align}
Here, $\phi(\bm r)$ and $n$ are the electric potential and electronic density; $\eta$ and $\rho$ are the viscosity coefficient and the electric resistivity, respectively. It is assumed that the velocity $\bm v(\bm r)$ of the electronic fluid is small, so the higher-order in $\bm v$ terms are dropped (see discussion in Ref. [\onlinecite{TorreGeim2015}]). In addition, the stationary continuity equation for current density $\bm j = ne \bm v$ is used:
\begin{align}
  0 =\bm\nabla \cdot \bm j(\bm r) = ne\bm \nabla \cdot \bm v(\bm r),
   \label{contin} 
\end{align}
where, in the last equality, we assumed that the electronic liquid is incompressible at hydrodynamic length scales, i.e.,  $n(\bm r) = \rm const$. 

\begin{figure}
	(a)\includegraphics[width=0.95\linewidth]{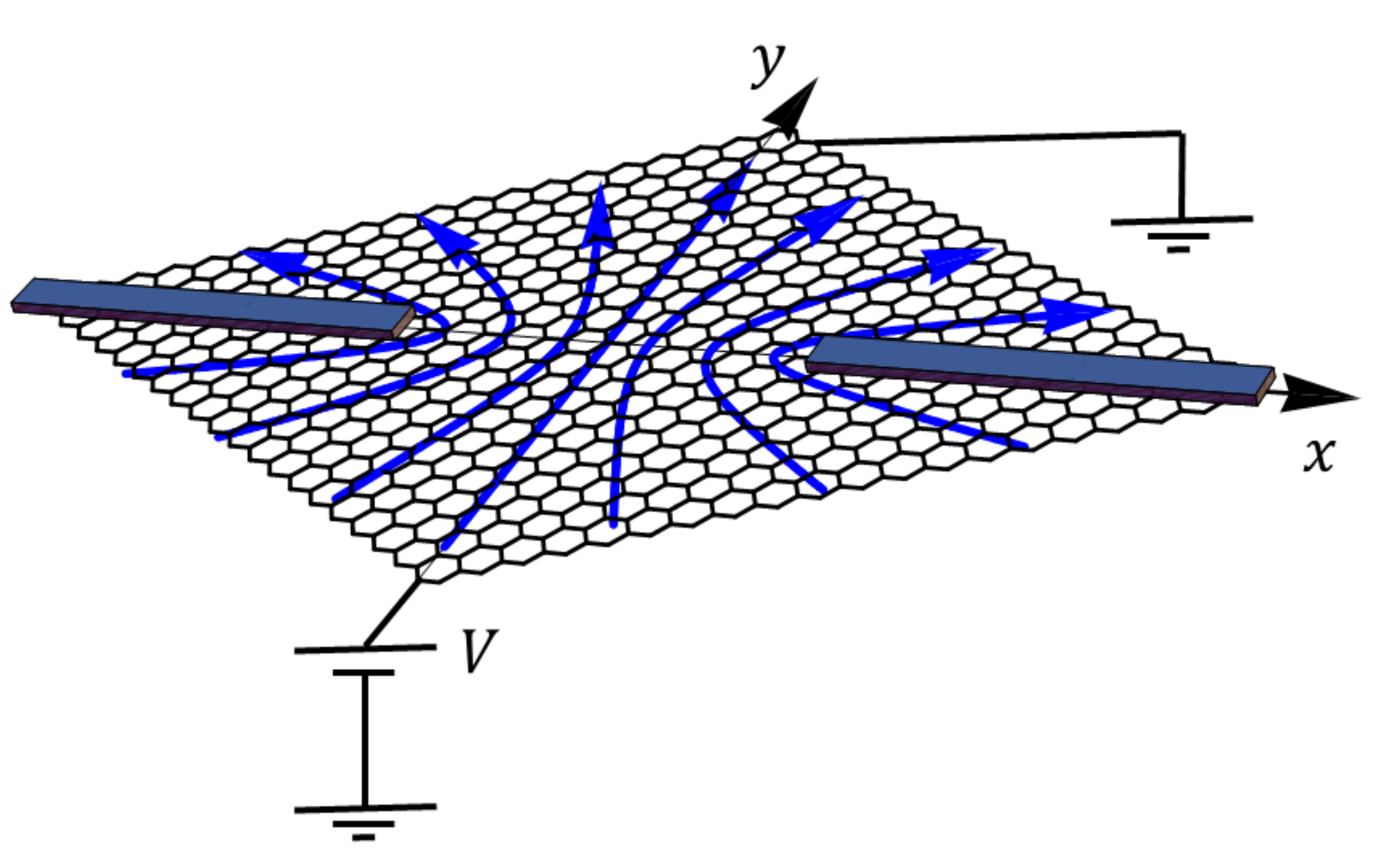}\\  \vspace{1cm} 
	(b)\includegraphics[width=0.95\linewidth]{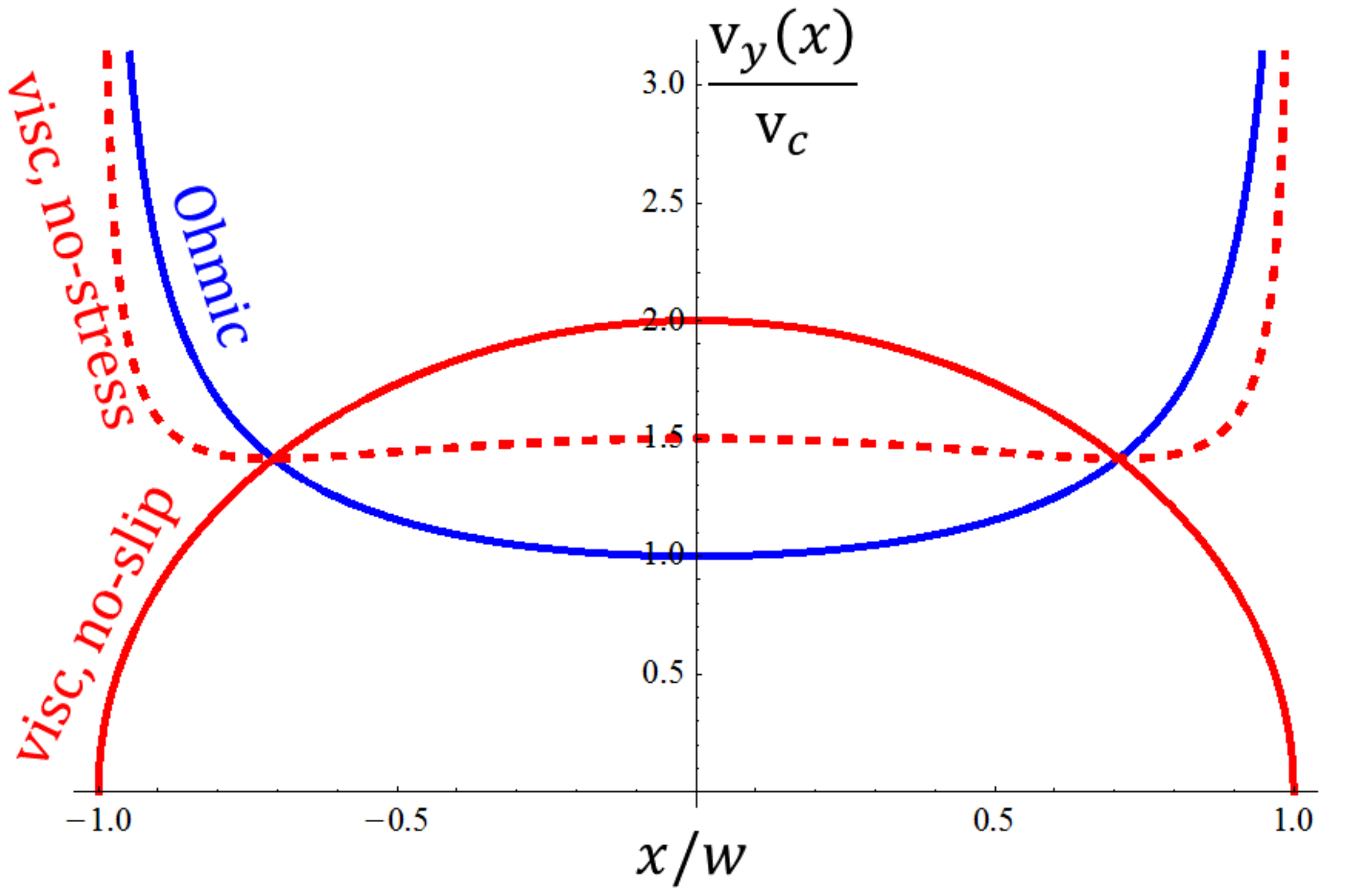}
	\caption{(a)  Schematic representation of the electric current flow in graphene through a slit of finite width $2w$. The scale of a graphene lattice is artificially enlarged for visualization. The distribution of current within the slit (i.e. at $|x|<w$ and $y=0$) may allow to distinguish between the viscous/non-viscous regimes as well as clarify the role of the boundary conditions.
	(b) The current distribution within the slit in the Ohmic  (\ref{ohmic_slit}), viscous no-slip (\ref{noslip_slit}), and viscous no-stress (\ref{nostress_slit}) cases. To plot them simultaneously, we set the common normalization constant $v_c = I/\pi ne w$, which corresponds to fixed total current $I$. } 
	\label{fig:3D_flow}
\end{figure}

We intend to solve Eqs.~(\ref{Stokes}) and (\ref{contin}) for the ``slit'' geometry. To be more specific, we assume that the graphene sheet is divided by the opaque (for electrons) barrier with a slit of finite width $2w$ as illustrated in Eq.~\ref{fig:3D_flow}(a). For the purposes of analytical calculations, we assume that the barrier is infinitely thin. 

The specifics of the boundary conditions imposed by the barrier is crucial for determining the profile of the flow. In the microscopic approach, the pioneering work by Fuchs \cite{Fuchs1938} discussed two types of boundary conditions for electrons: (i) the diffuse and (ii) specular scattering. In the phenomenological hydrodynamic approach, the  boundary conditions on each side of the impenetrable barrier may be formulated in a concise form, 
\begin{equation}
  \begin{aligned}
  & v_y|_{|x|>w,\,\,y\to 0} = 0, \\
  & v_x|_{|x|>w,\,\,y\to 0} = \lambda\,(\nabla_y v_x)|_{|x|>w,\,\,y\to 0}.
  \end{aligned} 
  \label{bc_viscous}
\end{equation}
The first of these two equations states that the normal component of the velocity  vanishes at the barrier. The second equation states that the tangential velocity at the boundary is proportional to the viscous stress. The parameter $\lambda$ allows to interpolate between the no-slip ($\lambda = 0$) and no-stress ($\lambda = \infty$) boundary conditions. There is no consensus in the literature (see Ref.  [\onlinecite{TorreGeim2015}] vs [\onlinecite{Guo3068}]) regarding which boundary condition is appropriate for graphene. Recent theoretical work~[\onlinecite{KiselevSchmalian2019}] discussed a relation between the microscopic and hydrodynamic boundary conditions. 

By inspecting the left-hand-side of Eq.~(\ref{Stokes}), it is instructive to define the parameter 
\begin{align}
l = \frac{1}{ne}\sqrt{\frac{\eta}{\rho}}, \label{l_param}
\end{align}
which has units of length. Comparison of $l$ with the geometric scale of the problem $w$ allows us to define the two regimes in which (i) the Ohmic term dominates ($l \ll w$), or (ii) the viscous term dominates ($l \gg w$). We discuss the current distribution in these limiting cases in the following Section. Then, in Sec.~\ref{sec:crossover}, we discuss the crossover between the two limits.

\section{Current distribution in the limiting cases.} \label{sec:limiting_cases}

\subsection{Current distribution in the Ohmic limit ($l/w \to 0$).} \label{sec:ohmic}
As a warm up, we consider the Ohmic limit $l/w \to 0$, in which we may drop the viscous ($\propto \eta$) term in Eq.~(\ref{Stokes}). In order to resolve the continuity Eq.~(\ref{contin}), we introduce the stream function $\bm v(\bm r) = [\hat {\bm z}\times \bm\nabla \psi(\bm r)]$. Then, Eq.~(\ref{Stokes}) reduces to
\begin{align}
[\hat {\bm z} \times \bm \nabla\psi(\bm r)] = -\frac{1}{ne\rho} \bm \nabla \phi(\bm r). \label{ohmslaw}
\end{align}
We seek a solution of Eq.~(\ref{ohmslaw}) with the normal component of velocity vanishing at the wall. That boundary condition amounts to $\psi$ being constant\footnote{Here we use that $\psi$ is defined up to a constant, so we may arbitrary shift it for our convenience.} at the two sides of the barrier, i.e. $\psi(\bm r)|_{x>w,y\to 0} = 0$ and $\psi(\bm r)|_{x<-w,y\to 0} = \psi_0$. The constant $\psi_0$ is related to the total current $I$ flowing through the slit, $\psi_0 = -I/ne$. 
Equation~(\ref{ohmslaw}) may be interpreted\cite{LL_hydro_book,FalkovichLevitovPRL2017} as the Cauchy-Riemann condition for an analytical function of a complex variable $z = x+iy$: 
\begin{align}
f(z) = \frac{\phi(\bm r)}{ne\rho} + i \,\psi(\bm r). \label{fz_ohmic}
\end{align}
Then, it is practical to perform a conformal transformation \footnote{It is instructive to view that conformal transformation as a sequence of two mappings, $z_1=g(f(z))$. The first one, $\tilde z = f(z) = \left(z + \sqrt{z^2-w^2}\right)/w$, is the inverse to the Joukowsky transform, and it maps the slit geometry onto the upper half-plane (i.e.,  $\tilde z = \tilde x + i \tilde y$ with $\tilde y > 0$). The second one, $z_1 = g(\tilde z) = \ln \tilde z$ transforms the upper-half plane into the horizontal stripe (i.e. $z_1 = x_1+i y_1$ with $\pi>y_1>0$). }
to a new variable $z_1 =  \ln \left[\left(z+\sqrt{z^2-w^2}\right)/w\right]$, in which the complicated slit geometry (see Fig.~\ref{fig:3D_flow}(a)) transforms into a horizontal stripe, i.e. $-\infty<x_1<\infty$ and $0<y_1<\pi$. In the latter geometry, the boundary conditions at the edges of the stripe become Im$\left.f(z_1)\right|_{z_1 \to x_1+i\pi} = \psi_0 = -I/ne$ and Im$\left.f(z_1)\right|_{z_1\to x_1+i 0} = 0$. It is straightforward to find the function satisfying that boundary condition: $f(z_1) = - z_1\,I/ne\pi $. So, in the original variable $z$, we have
\begin{align}
f(z) = -\frac{I}{\pi ne}\ln \left[\frac{z+\sqrt{z^2-w^2}}{w}\right]. \label{ohmic_solution}
\end{align}
The functions $\phi$ and $\psi$ may be read off from Eq.~(\ref{ohmic_solution}) using Eq.~(\ref{fz_ohmic}). Few comments about the solution~(\ref{ohmic_solution}) are in order. (i) The potential is logarithmically large $\phi(r) = ne\rho\, {\rm Re} f(z) \sim (I\rho/\pi)\ln \,r/w$ at $r \to \infty$. Physically, it corresponds to a logarithmically large resistance $R \sim (\rho/\pi) \ln L/w$, where $L$ is the size of the system. (ii)~Using that $\psi(\bm r) = {\rm Im}\,f(z)$ and definition of $\psi(\bm r)$, one may evaluate the velocity 
\begin{align}
\left( \begin{array}{c} v_x(\bm r) \\ v_y (\bm r) \end{array} \right) = \frac{I}{\pi ne}     \left( \begin{array}{c} {\rm Re} [ 1/\sqrt{z^2-w^2}] \\ -{\rm Im}[ 1/\sqrt{z^2-w^2}] \end{array} \right).
\end{align}
Within the slit, the flow has only the $\hat{y}$ component,
\begin{align}
v_y|^{\rm Ohmic}_{|x|<w,\,\,y \to 0} = \frac{v_c}{\sqrt{1-(x/w)^2}},\quad v_c = \frac{I}{\pi new}. \label{ohmic_slit}
\end{align}
Although the velocity has a square root divergence at the edges, the total current flowing through the slit is finite, and satisfies the current conservation law, $\int_{-w}^w dx\,nev_y(x) = \int_{-w}^w  dx\,\frac{nev_c}{1-(x/w)^2} = I$. 

\subsection{Current distribution in the viscous limit ($l/w \to \infty$).} \label{sec:viscous}
It was realized\cite{Guo3068} that the conductance in the viscous limit, i.e. at $l/w \to \infty$, with no-slip boundary conditions may surpass the ballistic limit. In this section, we complement the result of that study by considering the viscous limit with no-stress boundary conditions.  Although the conductance in the no-stress and no-slip cases behaves similarly, the velocity profiles differ significantly. The velocity vanishes at the edges of the slit in the no-slip case~\cite{Guo3068}. In contrast, the velocity profile in the no-stress case has a divergence similar to Eq.~(\ref{ohmic_slit}). 

In the viscous limit, the Ohmic term  ($\propto \rho$) may be dropped, and the Stokes equation~(\ref{Stokes}) becomes
\begin{align}
    \eta\, \Delta \,\bm v(\bm r) = ne \,\bm \nabla\phi(\bm r). \label{Stokes_no_ohm}
\end{align}
We follow Ref.~[\onlinecite{FalkovichLevitovPRL2017}] and introduce vorticity $\omega(\bm r) = {[\bm \nabla\times \bm v(\bm r)]_z}$, so  Eq.~(\ref{Stokes_no_ohm}) reduces to
\begin{align}
    [\hat {\bm z} \times \bm \nabla \omega(\bm r)] = \frac{ne}{\eta} \bm \nabla \phi(\bm r). \label{vorticity_potential}
\end{align}
We find velocity $\bm v(\bm r)$ in two steps: (i) first we solve Eq.~(\ref{vorticity_potential}), and (ii) next we compute $v(\bm r)$ from the evaluated $\omega(\bm r)$. 

(i) We proceed to solving the linear partial differential Eq.~(\ref{vorticity_potential}). We follow Ref.~[\onlinecite{FalkovichLevitovPRL2017}] and note that functions $\omega(\bm r)$ and $\phi(\bm r)$ satisfy the Cauchy-Riemann conditions for the analytical function of the complex variable $z = x+iy$:
\begin{equation}
   f(z) = -\frac{ne}{\eta}\phi(\bm r)+ i\,\omega(\bm r).
\label{analytical_fz}
\end{equation}
We intend to compute the function $f(z)$ in the upper half-plane, i.e. for $y>0$. For that, let us establish the boundary condition satisfied by $f(z)$ on the real axis, i.e, at $z \to x+i\,0$. It is convenient to set the electric potential $\phi(\bm r)$, such that $\phi(\bm r)|_{r\to \infty,\,{\pi>\varphi>0}} = 0$ and $\phi(\bm r)|_{r\to \infty,\,2\pi>\varphi>\pi} = V$, where $V$ is the applied bias and $\varphi$ is the polar angle of vector ${\bm r}$. Then, by invoking the symmetry of the problem, the electric potential is constant within the slit, i.e., $\left.\phi(\bm r)\right|_{|x|<w,y \to +0} = V/2$. Further, the no-stress boundary condition, i.e. setting $\lambda=\infty$ in Eq.~(\ref{bc_viscous}), renders the vorticity to vanish at the barrier, i.e., $\omega|_{|x|>w,\,y\to +0} = 0$. We may collect these boundary conditions in a concise way for the function $f(z)$ defined in Eq.~(\ref{analytical_fz}),
\begin{equation}
    \begin{aligned}
       & {\rm Re}\, f(z)|_{|x| < w,\, y \to +0} = -\frac{neV}{2\eta}, \\
       & {\rm Im}\,f(z)|_{|x| > w,\,y\to +0} = 0,\\
       & f(r e^{i\varphi})|_{r\to \infty,\,\,0<\varphi<\pi} = 0.
    \end{aligned} \label{mixed_boundary_condition}
\end{equation}
This is a mixed boundary value problem \cite{LavrentievShabatBook}. To solve it, we introduce an auxiliary complex function
\begin{align}
    \tilde f(z) = i f(z) \sqrt{z^2-w^2},
\end{align}
for which the boundary condition (\ref{mixed_boundary_condition}) transforms into ${\rm Re}\, \tilde f(z)|_{y\to +0} = \frac{neV}{2\eta}\sqrt{w^2-x^2}\,\,\theta(w-|x|)$.
Now, we may apply the Schwarz integral to the function $\tilde f(z) = \frac{1}{\pi i} \int_{-\infty}^\infty dx \frac{{\rm Re} \tilde f(x)}{x-z}$, evaluate that integral, 
and obtain the function
\begin{align}
     f(z) = \frac{neV}{2\eta}\left[-1+\frac{z}{\sqrt{z^2-w^2}}\right]. \label{fz1}
\end{align}

(ii) Now, we may compute the velocity from the evaluated vorticity $\omega(\bm r) = {\rm Im}\,f(z)$. It is convenient to switch to the independent variables $z = x+i y$ and $\bar z = x-iy$. The velocity  satisfies the continuity equation $\bm\nabla\cdot\bm v = 0$ and equation on vorticity $(\bm\nabla\times\bm v)_z = \omega(\bm r)$. The pair of these equations may be written in a compact form as  $\partial_{\bar z} (v_y+iv_x) = {\rm Im} f(z)$. That equation may be integrated by writing ${\rm Im} f(z) = \frac{1}{2i}\left[f(z)-f(\bar z)\right]$ and using the explicit expression for $f(z)$: 
\begin{equation}
    v_y+iv_x =  \frac{neV}{8i\eta}\left[\frac{z\bar z}{\sqrt{z^2-w^2}}-\sqrt{{\bar z}^2-w^2}+C(z)\right],
\end{equation}
where the function $C(z)$ is some analytical function of $z$.  
In order to determine $C(z)$, note that the velocity field is restricted by several constraints: (a) the component $v_x|_{|x|<w,y\to +0} = 0$ vanishes within the slit, (b) the component $v_y|_{|x|>w,y\to +0} = 0$ vanishes outside of the slit, and (c) $|\bm v|\propto\frac{1}{r}$ at large $r$. They prompt us to choose the following ansatz: $C(z) = A/\sqrt{z^2-w^2}$. The numerical constant $A$ may be determined by matching with the known behavior of the velocity field\footnote{From Eq.~(2) of Ref.~\onlinecite{FalkovichLevitovPRL2017}, we extract the large-$r$ behavior of velocity $\bm v(\bm r) = \frac{I\,\bm r}{\pi ner^2}\left( 1- \frac{1}{2}\cos2\theta\right)$. Here $I$ is the total current, $\theta$ is the polar angle.} at large $r$,  producing $A = -2w^2$. 
So, we may obtain the velocity within the slit
\begin{align}
    v_y|_{|x|<w,\,\,y \to 0}^{\rm visc,\,no-st} = v^{\rm no-st}_c \left[\sqrt{1-(x/w)^2}+\frac{1}{2\sqrt{1-(x/w)^2}}\right], \label{nostress_slit}
\end{align}
where $v^{\rm no-st}_c = {newV}/{4\eta}$. Evaluating the total current through the slit $I = ne\int_{-w}^{w} dx\,v_y$, we find the conductance $G^{\rm no-st} = I/V = \pi (new)^2/4\eta$. Let us contrast Eq.~(\ref{nostress_slit}) with the velocity distribution evaluated\cite{Guo3068} for the no-slip boundary condition 
\begin{align}
v_y|_{|x|<w,\,\,y \to 0}^{\rm visc,\,no-sl} = 2\,v^{\rm no-sl}_c \sqrt{1-(x/w)^2}, \label{noslip_slit}
\end{align}
where $v^{\rm no-sl}_c = {newV}/{8\eta}$. The conductance in the no-slip case is twice smaller, $G^{\rm no-sl} = G^{\rm no-st}/2$.

\subsection{Comparison between the Ohmic and viscous limits}
We summarize the results of the current section by plotting the velocity distributions in the Ohmic~(\ref{ohmic_slit}), viscous no-stress (\ref{nostress_slit}), and viscous no-slip (\ref{noslip_slit}) limits in Fig.~\ref{fig:3D_flow}(b). Observe that both the Ohmic~(\ref{ohmic_slit}) and viscous no-stress~(\ref{nostress_slit}) distributions have an integrable $v_{y}\propto 1/\sqrt{x\pm w}$ singularity at the edges of the slit. Physically, that divergence stems from the requirement to accommodate the non-vanishing flow along the impenetrable boundary. The profiles of velocity for the Ohmic~(\ref{ohmic_slit}) and viscous no-stress (\ref{nostress_slit}) limits appear similar qualitatively. Therefore, it would be challenging to experimentally distinguish the two limits. 

In contrast, the velocity profile ~(\ref{noslip_slit}) in the case of the no-slip boundary conditions is a convex function with a maximum at the center of the interval~$(-w,w)$. It is significantly different from the concave velocity profile in case of the Ohmic flow. 
Once the Ohmic ($\propto \rho$) and viscous ($\propto \eta$) terms become of comparable strength, i.e., $l/w \sim 1$, the solutions~(\ref{ohmic_slit}) and (\ref{noslip_slit}) corresponding to the limiting cases are not applicable, and we expect a crossover between the concave and convex velocity distributions across the slit ($|x|\leq w$). In the next section, we develop a method of integral equation to describe that crossover.

\section{Crossover between the Ohmic and no-slip viscous limits ($l/w\sim 1$).} \label{sec:crossover}


\subsection{Integral equation}

In the spirit of  Refs.~[\onlinecite{FalkovichLevitovNatPhys2016},\onlinecite{FalkovichLevitovPRL2017}], we find the solution of the ``point-source'' (ps) problem
\begin{align}
   \psi_{\rm ps}(x,y) = \int_{-\infty}^\infty \frac{dk_x}{2\pi i\,k_x}\,&\frac{e^{ik_xx}}{q-|k_x|} \left[q\, e^{-y\,|k_x|} - |k_x|\,e^{-y\, q}\right],\nonumber\\
   & \qquad\quad q = \sqrt{k_x^2+l^{-2}}, \label{psips}
\end{align}
where the parameter $l$, defined in Eq.~(\ref{l_param}), measures the relative strength of the viscous and Ohmic terms. Equation~(\ref{psips}) solves Eqs.~(\ref{Stokes}) and (\ref{contin}) for arbitrary $\eta$ and $\rho$ with no-slip boundary condition 
and a ``point-source'' current at the boundary $y=0$. In other words, it satisfies $v_x|_{y\to +0}=-\nabla_y \psi_{\rm ps}|_{y\to +0} = 0$ and $v_y|_{y\to +0}=\nabla_x \psi_{\rm ps}|_{y\to +0} = \delta(x)$. One may view Eq.~(\ref{psips}) as a Green's function allowing to relate $\psi(x,y)$ in the plane to the velocity $v(x)$ within a finite-width slit:
\begin{align}
\psi(x,y) = \int_{-w}^w\,dx'\,\psi_{\rm ps}(x-x',y)\,v(x'). \label{integral_psi}
\end{align}
For clarity, the components $v_x$ and $v_y$ stand for the velocity at arbitrary $\bm r$, whereas $v(x) \equiv v_y(x,y)|_{y\to 0}$ denotes the velocity distribution within the slit. Naturally, $\psi(x,y)$ satisfies the correct boundary conditions at $y = 0$ as well as the condition on the total current at $r \to \infty$. In addition, the velocity distribution must satisfy the  symmetry condition that $y=0$ is the inflection point for $v_x$, which amounts to $\left.\nabla_y^3 \psi\,\right|_{|x|<w,y \to +0} =0$ in terms of the stream function $\psi$. Substituting Eq.~(\ref{integral_psi}) in the latter symmetry condition \footnote{To be accurate, we multiply by $l^2$, i.e. $-l^2\left.\nabla_y^3 \psi\,\right|_{|x|<w,y \to +0} = 0$ corresponds to Eq.~(\ref{kernel}).} and massaging it yields an integral equation on the unknown velocity profile $v(x)$
\begin{align}
    &\dashint_{-w}^{w} dx'\, K(x-x')\, v(x') = 0, \label{integral_eq} \\
    &  K(x) = \lim_{\delta \to +0}\int_0^\infty \frac{dt\,e^{-t\,\delta }}{l}\,\sin\left(t\, \frac{x}{l}\right)\,\frac{\sqrt{t^2+1}}{\sqrt{t^2+1}-t}. \label{kernel} 
\end{align}
So the problem reduces to finding a null vector of the integral operator with kernel $K(x)$. In addition, we impose a boundary condition $v(\pm w) = 0$. To ensure convergence, the integrand in Eq.~(\ref{kernel}) contains\footnote{The exponential term $e^{-t\,\delta}$ in Eq.~(\ref{kernel}) is a remnant of the terms $e^{-|k_x|y}$ and $e^{-q y}$ in Eq.~(\ref{psips})} an exponentially decaying term $e^{-t\,\delta}$. In the absence of that term, the integrand diverges at large $t$, which represents the singularity of the kernel at $x\to 0$. In order to expose that singularity, we re-write the rational function of the integrand in Eq.~(\ref{kernel}) as:
\begin{align*}
    &K(x) = \\
    &\lim_{\delta\to 0}\int_0^\infty \frac{dt\, e^{-t\,\delta}}{l}\,\sin\left(t\, \frac{x}{l}\right)\,\left[2\,t^2+\frac{3}{2} -\frac{1}{2(t+\sqrt{t^2+1})^2}\right].
\end{align*}
We may explicitly evaluate the integrals corresponding to the first two terms in the square brackets and retain the last term in $K_{reg}(x)$:
\begin{align}
   K(x) &= -\frac{4\,l^2}{x^3}+\frac{3}{2x}+K_{reg}(x), \label{kernel1}
   \\ & K_{reg}(x) = -\int_0^\infty \frac{dt}{2\,l}\,\frac{\sin\left(t\, (x/l)\right)}{(t+\sqrt{t^2+1})^2}.\nonumber
\end{align}
The first two terms in Eq.~(\ref{kernel1}) are singular, and, correspondingly, the integral~(\ref{integral_eq}) is understood in the sense of Cauchy's principal value. In contrast, the integral in $K_{reg}(x)$ converges well and, so, the regularizing exponent is dropped. It has the following asymptotes: $K_{reg}(x) = (x/l^2) \ln (l/|x|)$ and $K_{reg}(x) = -1/2x+2\,l^2/x^3+\mathcal O( l^4/x^5)$ at $x/l \ll 1$ and $x/l\gg 1$, respectively.

\subsection{Limiting cases}
Let us demonstrate that the limiting cases are consistent with the integral equation approach. First, consider the Ohmic limit $l \to 0$, in which case the kernel (\ref{kernel1}) becomes $K(x) = \frac{1}{x}$. Then, it is straightforward to check that the Ohmic velocity profile~(\ref{ohmic_slit}) satisfies the integral equation~(\ref{integral_eq}): 
\begin{align}
   \dashint_{-w}^wdx'\,\frac{1}{x-x'}\,\left[\frac{v_c}{\sqrt{1-x'^2}}\right] = 0.
\end{align}
In the opposite strongly viscous case $l \to \infty$, the kernel behaves as $K(x) = -4l^2/x^3$. One may show that the velocity profile~(\ref{noslip_slit}) satisfies the corresponding integral equation~(\ref{integral_eq}),
\begin{align}
   \dashint_{-w}^wdx'\,\frac{-4l^2}{(x-x')^3}\,\left[2v_c\sqrt{1-x'^2}\right] = 0,
\end{align}
and the boundary condition $v(\pm w) = 0$ at the edges of the slit. The consideration above prompts the following interpretation of the singular terms in kernel~(\ref{kernel1}). The two terms $\propto 1/x$ and $\propto l^2/x^3$ correspond to the Ohmic and viscous parts of the kernel, respectively.

\subsection{Numerical solution}
\begin{figure}
	\includegraphics[width=0.95\linewidth]{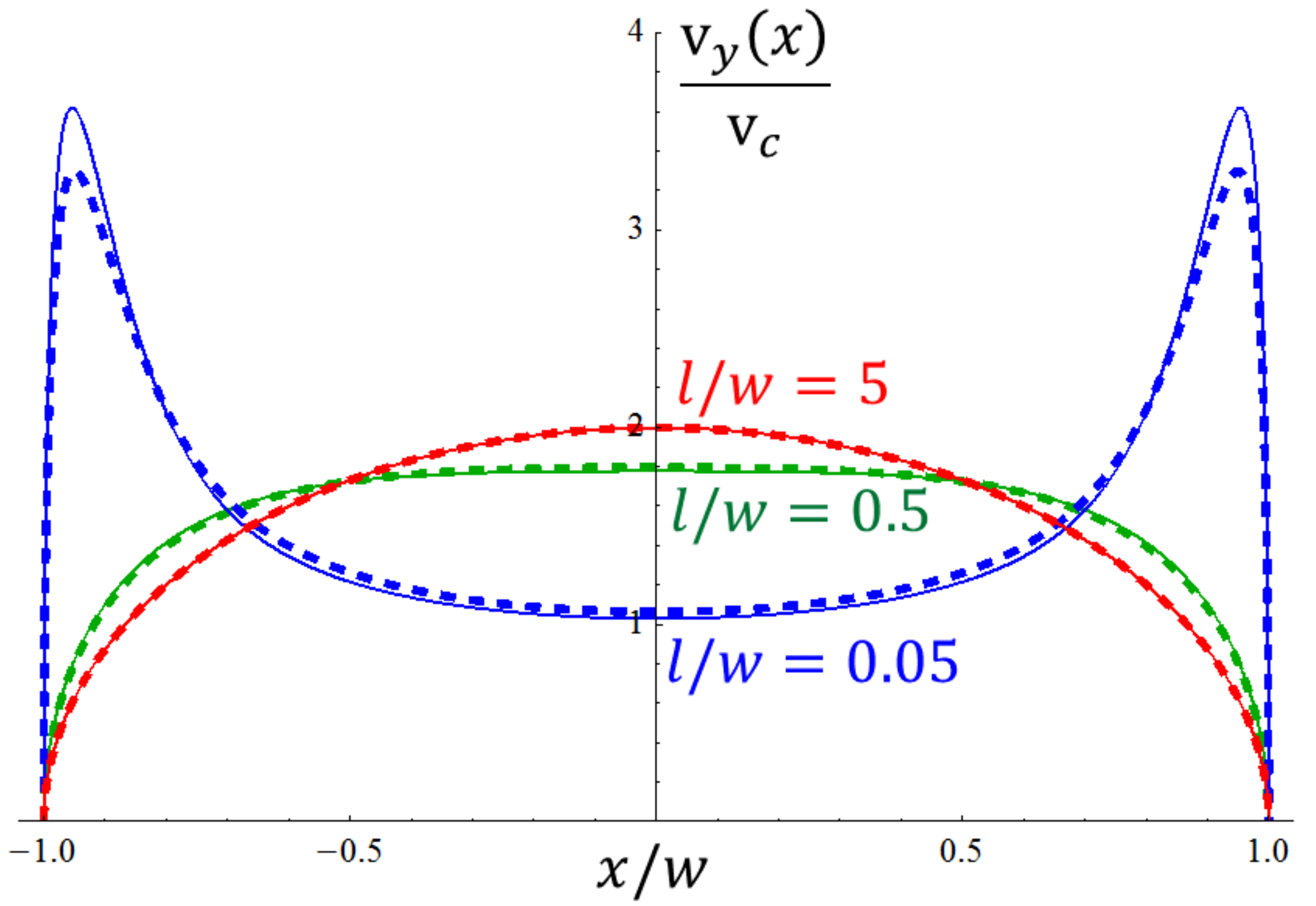}
	\caption{Normalized velocity profile through the slit evaluated for the no-slip boundary condition. We present results ranging from the strongly viscous $l/w \gg 1$ to strongly Ohmic $l/w \ll 1$ regimes. The crossover between the two regimes occurs at the intermediate $l/w \simeq 0.5$. The numerical and (approximate) analytical~(\ref{analytical_vx}) curves are shown with solid and dashed lines, respectively.} 
	\label{fig:crossover}
\end{figure}
Equation~(\ref{kernel1}) is conveniently split in singular~($\propto 1/x$ and $\propto l^2/x^3$) as well as non-singular $K_{reg}(x)$ terms. The strategy is to simplify the singular terms by analytical methods, whereas the non-singular term may be treated numerically. 

We proceed by substituting the kernel~(\ref{kernel1}) in Eq.~(\ref{integral_eq}) and recognize that the viscous term ($\propto l^2/x^3$) may be written via a second derivative:
\begin{align}
    &-2l^2 \frac{d^2}{dx^2}\left[\dashint_{-w}^w dx' \frac{v(x')}{x-x'}\right] + \frac32\left[\dashint_{-w}^w dx' \frac{v(x')}{x-x'}\right] \nonumber\\
    &\qquad\qquad\qquad+ \int_{-w}^w dx'\, K_{reg} (x-x')\,v(x')=0. \label{integro_differential_eq}
\end{align}
In order to tackle this integro-differential equation, we employ the Chebyshev polynomials of both first $T_n(x)$ and second $U_n(x)$ kinds.\footnote{{The Chebyshev polynomials of the first and second kinds are defined as $T_n(\cos\theta) = \cos(n\,\theta)$ and $U_n(\cos\theta) = \sin[(n+1)\theta]/\sin\theta$, respectively.
}} They are tailored for a problem on a finite interval. We expand the velocity profile in series  
\begin{align}
& v(x) =  \frac{v_c}{\sqrt{1-(x/w)^2}}\sum_{n=0}^\infty c_n\,T_{2n}(x/w),  \label{chebyshev_expansion}
\end{align}
where $v_c = \frac{I}{\pi new}$ denotes the characteristic value of velocity. The summation is carried over the polynomials of even order, which are even functions of $x$, thus corresponding to the symmetry of the problem. The value of the first coefficient $c_0=1$ is fixed by the constraint $\int_{-w}^w dx\, v(x) = I/ne$, whereas $c_n$ are unknown for $n\ge 1$.

The expansion~(\ref{chebyshev_expansion}) enables to rewrite Eq.~(\ref{integro_differential_eq}) as a system of linear equations, which may be solved numerically. Let us briefly sketch that procedure; the details are given in Appendix. Substituting the expansion~(\ref{chebyshev_expansion}) in the principal value integral appearing in Eq.~(\ref{integro_differential_eq}) yields
\begin{align}
\dashint_{-w}^w dx'\,\frac{v(x')}{x-x'} = - v_c \sum_{n=1}^{\infty} c_n\,\pi\, U_{2n-1}(x/w), \label{useful_identity}
\end{align}
where we used Eq.~(18.17.42) of Ref.~[\onlinecite{NIST_DLMF}]. The last term in Eq.~(\ref{integro_differential_eq}) may also be presented as a linear combination of $U_n(x)$ [see Eq.~(\ref{kreg})]. Therefore, by relying on the orthogonality of the polynomials $U_n(x)$, Eq.~(\ref{integro_differential_eq}) reduces to an infinite system of linear equations on the coefficients $(c_1,c_2,c_3,\ldots)$ [see Eq.~(\ref{linear_eq1})]. 
In addition, given Eq.~(\ref{chebyshev_expansion}) and the property $T_{2n}(\pm 1) = 1$, the boundary condition $v(\pm w)=0$ leads to the condition $\sum_{n\geq 1} c_n = -c_0 = -1$ [see Eq.~(\ref{linear_eq2})]. Truncating the matrix of that linear system, i.e., setting $c_n = 0$ for $n>N$, renders a finite system of linear equations
amenable to a numerical solution. The elements of that matrix depend on the parameter $l/w$, allowing us to investigate the crossover between the Ohmic and viscous flows. The evaluated coefficients $c_n$ are then substituted in Eq.~(\ref{chebyshev_expansion}) thereby producing the velocity profile.   

In Fig.~\ref{fig:crossover}, we present the result of the numerical procedure outlined above for the parameters ranging from the strongly viscous $l/w \gg 1$ to strongly Ohmic $l/w \ll 1$ regimes. In the latter regime $l/w \gg 1$, the velocity profile is a convex function with a single maximum at $x=0$. With decrease of $l/w$ (i.e. with the decrease of $\eta$), the profile further flattens at the center until the second derivative of velocity vanishes at $x=0$ for some critical value of parameter $l/w \simeq 0.5$. The two shallow maxima 
appear in the vicinity of $x=0$ for $l/w < 0.5$. With further decrease of $l/w$, the two maxima sharpen and drift towards the edges of the slit as the velocity profile approaches Eq.~(\ref{ohmic_slit}) evaluated in the Ohmic limit.

\subsection{Analytical interpolation between the viscous and Ohmic limits}
We recall that the distribution of the velocity $v(x)$ in the two limits can be obtained from an integral equation with the kernel truncated to the corresponding singular term [see Eqs. (24) and (25)]. Next, we note that the boundary values $v(-w)=v(w)=0$ would be enforced by the stronger singularity of the viscous $-4l^2/x^3$ part of the kernel (23) at any $l$, even if $l\ll w$ and the Ohmic term dominates everywhere except the vicinity of the ends of the slit. Therefore, it is clear that the qualitative behavior of $v(y)$ should be captured by a solution of the integral equation Eq. (26) with an omitted part $K_{\rm reg}$. The resulting equation,
\begin{align}
&-2l^2 \frac{d^2}{dx^2}\left[\dashint_{-w}^w dx' \frac{v(x')}{x-x'}\right] + \frac32\left[\dashint_{-w}^w dx' \frac{v(x')}{x-x'}\right] = 0\,
\label{differential_eq1}
\end{align}
can be solved analytically. Remarkably, this solution provides one with an excellent fit to the numerical results in a broad range of the ratios $w/l$ which includes the crossover between the concave and convex profiles of $v(x)$.

We view Eq.~(\ref{differential_eq1}) as a second-order differential equation. When solving it, we pick the odd in $x$ solution,
\begin{align}
\dashint_{-w}^w dx' \frac{v(x')}{x-x'} = C \sinh\left(\frac{x\sqrt 3}{2l}\right), \label{int_eq_ap}
\end{align}
where the constant $C$ will be determined below. In order to invert Eq.~(\ref{int_eq_ap}), we expand both the left- and right-hand sides of Eq.~(\ref{int_eq_ap}) in Chebyshev polynomials $U_{n}(x)$. For the left-hand side, we use Eq.~(\ref{useful_identity}). For the right-hand side, we evaluate an expansion 
\begin{align}
\sinh\left(\frac{\sqrt 3\, x}{2\,l}\right) = \frac{8\,l}{\sqrt{3}\,w} \sum_{n=1}^\infty\,n\, I_{2n} \left(\frac{\sqrt 3\, w}{2 \,l}\right) U_{2n-1}\left(\frac{x}{w}\right), \label{exp_U_ap}
\end{align}
where $I_{n}(x)$ are the modified Bessel functions. Thereby, the left- and right-hand sides of Eq.~(\ref{int_eq_ap}) are presented as series in orthogonal $U_{2n-1}(x)$ polynomials. So, the expansion coefficients $c_n$ may be read off: $c_n = - C\, n \,I_{2n} \left(\frac{\sqrt 3\, w}{2 \,l}\right)$ for $n>0$. Recall that the coefficient $c_0=1$ is determined by fixing the total current. So, we obtain the analytical expression for velocity
\begin{align}
& v(x) = \label{analytical_vx}  \\
&\frac{v_c}{\sqrt{1-(x/w)^2}}\left[1- C \sum_{n=1}^\infty n \,I_{2n} \left(\frac{\sqrt 3\, w}{2 \,l}\right)\,T_{2n}\left(\frac{x}{w}\right)\right]. \nonumber
\end{align}
The remaining constant $C$ is determined from the boundary condition $v(\pm w) = 0$, producing 
\begin{align}
C^{-1} = \sum_{n=1}^\infty n\,I_{2n}\left(\frac{w\sqrt 3}{2\,l}\right). \label{analytical_vxC}
\end{align}
For comparison, we superpose the numerical curves with analytical result~(\ref{analytical_vx}) in Fig.~\ref{fig:crossover}. As expected, the analytical and numerical curves agree perfectly at $l/w \gg 1$, where the viscous term in the kernel is dominant in the entire range $|x|\leq w$. It is remarkable that at $l/w \simeq 1$ and even at $l/w \ll 1$, the analytical curves give a very good approximation to the numerical results in that entire range. Our rationalization of such a good agreement that it is the competition between the singular terms in the kernel ($\propto l^2/x^3$ and $\propto 1/x$) that determines the velocity profile $v(x)$ through the slit. The regular term $K_{reg}(x)$ is subdominant and may only slightly renormalize the relative strength of the singular terms. Therefore the extrapolation by means of Eqs. ~(\ref{analytical_vx}) and (\ref{analytical_vxC}) provides a convenient way for a quantitative comparison of experimental results with theory predictions.
\section{Conditions for experimental observation of the Ohmic-to-viscous flow crossover}
\label{Sec:parameterdomain}

\begin{figure}
	\includegraphics[width=0.95\linewidth]{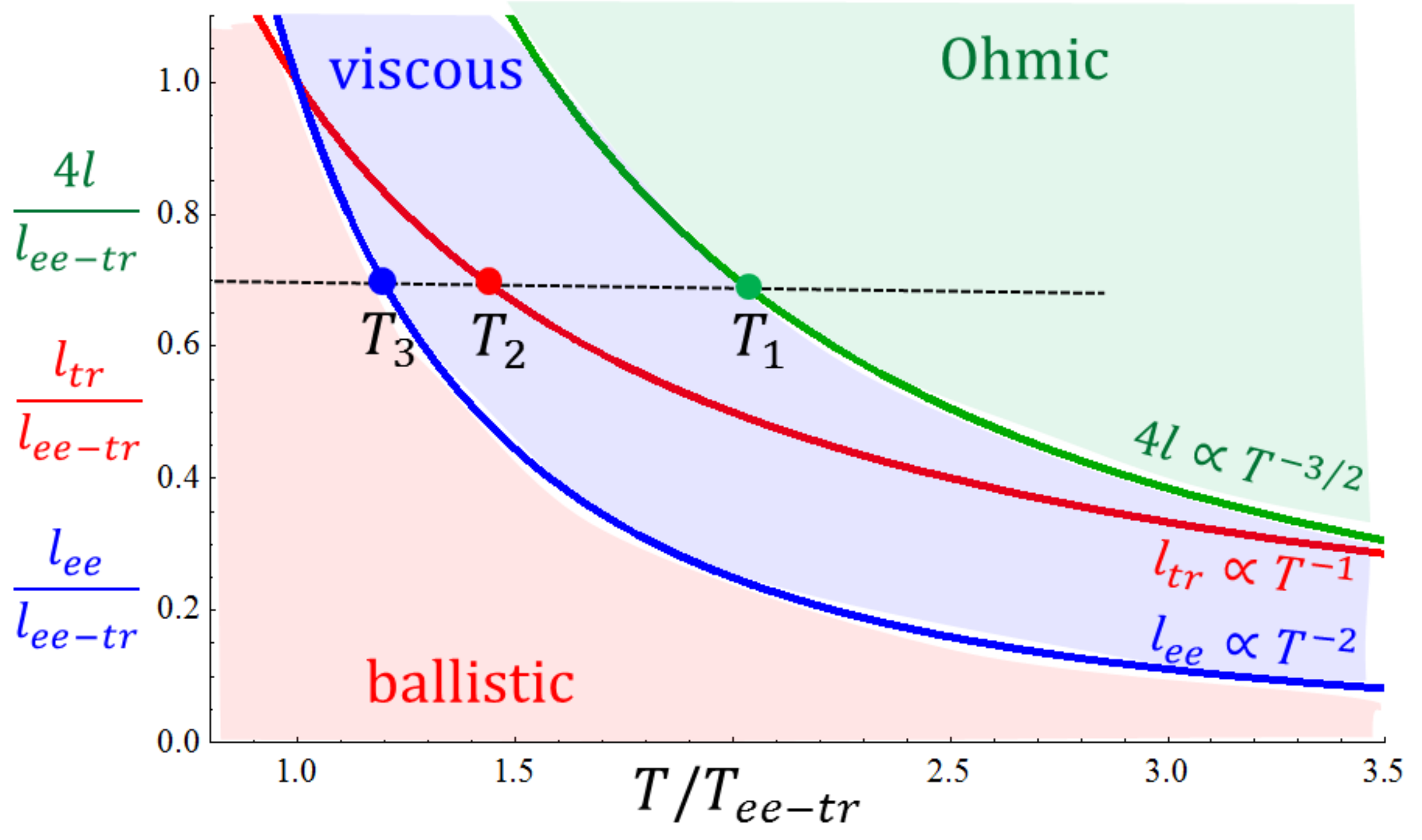}
	\caption{Diagram of different transport regimes in the $(T,l)$ plane. The lengths and temperature are normalized by values given in Eq.~(\ref{Teetr}). The lines corresponding to the transport mean free path $l_{tr}$ (red) and the mean-free path for the electron-electron  scattering $l_{ee}$ (blue), have distinct scaling with temperature [see Eq.~(\ref{leetr})].   Their geometric mean, shown in green, determines the Ohmic-to-viscous crossover line [see Eqs.~(\ref{lltree}) and (\ref{wl})].
	Lowering of a temperature at fixed electron density corresponds to a motion along some horizontal (dashed) line with vertical coordinate representing the fixed slit width $2w$. Its intersection with the three curves determines three temperatures: $T_1$, $T_2$, and $T_3$. At $T>T_1$, the flow through the slit is in the Ohmic regime. At $T=T_1$, the crossover to the viscous regime, discussed in this work,  occurs. At $T<T_2$, the notion of local conductivity becomes inapplicable, but the viscous flow regime persists; at point $T_3$, the viscous-to-ballistic crossover occurs~[\onlinecite{Guo3068}].} 
	\label{fig:l_vs_T}
\end{figure}

In experimental setting, the slit width $2w$ is fixed within a specific device. One may examine the effect of temperature $T$ and electron density $n$ variation  on the current density distribution within a slit. In this section, we address two questions which arise in that context: (i) what is the optimal width $2w$ for the observation of crossover, and (ii) what are the temperature and electron density at which the crossover is likely to occur. Apart from technological constraints limiting the long-scale homogeneity of a sample, additional considerations for choosing $w$ come from a remarkably long electron transport mean-free path $l_{tr}$ at low temperatures\cite{CoryDean2013}. The temperature dependence $l_{tr}(T)$ comes from the electron scattering off phonons. Upon lowering the temperature, the increase of $l_{tr}$ saturates at some value $l_{tr}(0) \sim 10\,\mu$m due to the residual scattering off impurities \cite{CoryDean2013}. 

The sample homogeneity requirement favors smaller values of $w$, so in the following we assume $w\ll l_{tr}(0)$ and account only for the phonon contribution to $l_{tr}$. Furthermore, considering the temperature dependence of $l_{tr}$, we focus on $T$ above the Bloch-Gr\"uneisen temperature~\cite{HwangPRB},
\begin{align}
    l_{tr}(T) = \frac{4\hbar^2 v_F^2 v_{ph}^2\,\rho_M}{\sqrt{\pi} D^2}\,\frac{1}{T\sqrt{n}}\,. \label{ltr}
\end{align}
Here $\rho_M$, $v_{ph}$, and $D$ are, respectively, the mass density, phonon velocity, and deformation potential in graphene, and $v_F$ is the Fermi velocity of the charge carriers; hereinafter $T$ is measured in units of energy. The viscosity is proportional to the electron mean free path $l_{ee}$ with respect to the electron-electron scattering~\cite{Guo3068}: $\eta=\nu nm=(1/4)v_Fl_{ee}nm$; here $n$ is the charge carriers density, and $m=p_F/v_F$ is the mass conventionally related to the Fermi momentum $p_F$ and velocity $v_F$ (for reference, we also introduced here the kinematic viscosity $\nu$ used instead of $\eta$ in some works~\cite{Bandurin2016}). The mean free path $l_{ee}=\alpha \hbar v_F^2p_F/T^2$ is also temperature-dependent. We may re-write $l_{ee}$ in terms of $n$ instead of $p_F$,
\begin{equation}
    l_{ee}(T)=\sqrt{\pi}\alpha\hbar^2\frac{v_F^2}{T^2}\sqrt{n}\,;
    \label{lee}
\end{equation}
the interaction constant $\alpha=e^2/(\hbar v_F\epsilon)$ depends on the dielectric constant $\epsilon$ of the environment (in re-writing, we accounted for the valley and spin degeneracy).
It is convenient to parametrize $l_{tr}(T)$ and $l_{ee}(T)$ by temperature $T_{ee-tr}(n)$ at which the two lengths equal each other, $l_{tr}(T_{ee-tr})=l_{ee}(T_{ee-tr})\equiv l_{ee-tr}(n)$, and by that length ($l_{ee-tr}$):
\begin{equation}
    T_{ee-tr}(n)=\frac{\pi\alpha}{4}\frac{D^2}{\rho_m v_{ph}^2}n\,;\,\,\, 
    l_{ee-tr}(n)=\frac{\sqrt{\pi}\alpha\hbar^2 v_F^2}{T^2_{ee-tr}(n)}\sqrt{n}.
    \label{Teetr}
\end{equation}
With these notations, we find
\begin{equation}
    l_{tr}(T)=l_{ee-tr}\frac{T_{ee-tr}}{T}\,;\quad 
    l_{ee}(T)=l_{ee-tr}\left(\frac{T_{ee-tr}}{T}\right)^2.
    \label{leetr}
\end{equation}
The temperature-dependent scattering lengths $l_{\rm tr}(T)$ and $l_{ee}(T)$ are plotted in Fig.~\ref{fig:l_vs_T} in units defined by Eq.~(\ref{Teetr}).

As shown in Sec.~\ref{sec:crossover}, the competition between the viscous and Ohmic terms defines the width $l$ of the boundary layer for the spatial distribution of the current density [see Eq.~(\ref{l_param})]. Using the Drude formula for resistivity, $\rho=mv_F/(ne^2l_{\rm tr})$, and the expression for viscosity, $\eta = (1/4)v_Fl_{ee}nm$, we may conveniently express $l$ in terms of $l_{tr}(T)$ and $l_{ee}(T)$:
\begin{equation}
    l=\frac{1}{2}\sqrt{l_{tr}(T)l_{ee}(T)}=\frac{1}{2}\,l_{ee-tr}
    \left(\frac{T_{\rm ee-tr}}{T}\right)^{3/2}
    .
    \label{lltree}
\end{equation}
For the current flow through a slit, the applicability of the hydrodynamic description requires that the width of the slit exceeds the electron-electron scattering length, i.e. $2w\gtrsim l_{ee}$, while using the notion of resistivity relies on $2w\gtrsim l_{tr}$. Under these conditions, we found the Ohmic-to-viscous crossover to occur at $w\approx 2l$. We rewrite this condition using  Eq.~(\ref{lltree}) as
\begin{equation}
    2w=2\,l_{ee-tr}
    \left(\frac{T_{ee-tr}}{T}\right)^{3/2}.
    \label{wl}
\end{equation}
Here, we multiply by 2 the left- and right-hand sides of Eq.~(\ref{wl}) in order to display it on par with $l_{ee}$ and $l_{tr}$ in Fig.~\ref{fig:l_vs_T}.  

Figure~\ref{fig:l_vs_T} sets the stage for determining the range of the slit widths $2w$ most favorable for observing the viscous flow, and the temperature of the Ohmic-to-viscous crossover at a given value of $2w$.  At $2w>(1/4)l_{ee-tr}(n)$, the crossover from  Ohmic regime to viscous flow occurs when the slit width $2w$ exceeds the mean free paths $l_{tr}$ and $l_{ee}$, justifying the hydrodynamic description of electron liquid. This type of crossover is considered in detail in this work. One may see from Fig.~\ref{fig:l_vs_T} that a slit of width $2w\lesssim l_{ee-tr}(n)$ is the most favorable for observing this type of crossover. Further reduction of temperature makes scattering off phonons irrelevant, once $l_{tr}$ exceeds the slit width. At even lower temperatures, the viscous flow gives way to ballistic electron propagation~\cite{Guo3068}.

The temperature of the Ohmic-to-viscous crossover increases with the decrease of $2w$. At $2w=(1/4)\,l_{ee-tr}(n)$ the crossover temperature is $4T_{ee-tr}$, see Eq.~(\ref{wl}). (The corresponding point is slightly off the plot in Fig.~\ref{fig:l_vs_T}.)
At $2w<(1/4)\,l_{ee-tr}(n)$ the crossover to viscous flow occurs upon lowering the temperature, once $l_{tr}(T)$ exceeds the slit width. This type of crossover is not considered in this work; however, it is clear that the concave-to-convex transition would occur in the case of no-slip boundary conditions, while the current flow profile would remain concave in the case of no-stress boundary condition [cf. Eqs.~(\ref{nostress_slit}) and (\ref{noslip_slit})].

The temperature domain for the viscous flow is also constrained from below (see Fig.~\ref{fig:l_vs_T}): the charge carrier transport enters the ballistic regime once both $l_{ee}(T)$ and $l_{tr}(T)$ exceed $2w$. Neglecting the electron diffraction, which occurs on the length scale of the Fermi wavelength $2\pi\hbar/p_F$, one finds a flat distribution ($v_y|_{|x|<w,\,\,y \to 0}$ independent of $x$) for the ballistic flow. We note here that our numerical solution  for the velocity profile in the vicinity to the Ohmic-to-viscous crossover also shows quite flat distribution (see the profile for $l/w=0.5$ in Fig.~\ref{fig:crossover}). One needs a resolution better than $0.1w$ to see the rounding of the profile near the slit ends, indicative of the viscous flow.

Using the parameters for graphene~\cite{Young2020} ($v_F = 10^6$\,m/s, $v_{ph} = 2.1\times 10^{4}$\,m/s, $D = 25$\,eV, $\alpha \approx 1$), we estimate $T_{ee-tr} = 27\, (n/n_0)$\,K and $l_{ee-tr} = 13 \, (n_0/n)^{3/2}\,\mu$m. Here $n_0 = 10^{12}\, \rm cm^{-2}$ is a typical density achieved in experiments~\cite{Bandurin2016,Young2020}. 
We note that $T_{ee-tr}=27$~K at $n = 10^{12}\, \rm cm^{-2}$ falls in the middle between the high-temperature ($\propto 1/T$) and low-temperature ($\propto 1/T^4$) asymptotes for $l_{tr}$ which is limited by electron-phonon scattering~\cite{HwangPRB}; in this case $T_{ee-tr}$ should be viewed merely as a scale for measuring $T$ (this is why we use a dashed line for a part of the $l_{tr}(T)$ curve in Fig.~\ref{fig:l_vs_T}). Equations~(\ref{ltr})-(\ref{wl}) assume that the electron thermal energy is small compared to the Fermi energy $E_F$; this condition is easily satisfied, as $E_F=116$~meV at $n = 10^{12}\, \rm cm^{-2}$. The corresponding Fermi wavelength, which defines the scale for the electron diffraction at the slit edges, is fairly small at approximately $3.5\times 10^{-6}$~cm. 
According to our estimates, the lowest temperature $T=128$~K in the experiment~\cite{Young2020} at density $n= 10^{12}\, \rm cm^{-2}$ and slit width of $4\,\mu{\rm m}$ was fairly close to the point of crossover between the Ohmic and viscous flows.

\section{Conclusion} \label{sec:conclusion}
The goal of this work is to identify the favorable conditions for observing the viscous electron flow in graphene and to facilitate an accurate measurement of the density profile of the current constrained by the device geometry. We find the slit geometry promising as it creates large gradients of electric potential and rapid spatial variations of electron velocity near the edges of the wall cut by the slit. It may help gaining information about the boundary conditions for the electron flow from the local-probe measurements~\cite{Young2020,ShahalIlani2019,Walsworth2019}.

In the case of Ohmic flow, the divergent electric field causes $1/\sqrt{x}$ singularities of the current density at the edges of the slit [see Eq.~(\ref{ohmic_slit})]. We establish that the velocity in the viscous flow with no-stress boundary condition also results in $1/\sqrt{x}$ divergence at the edges [see Eq.~(\ref{nostress_slit})]. It qualitatively resembles the velocity profile in the Ohmic limit, making it difficult to distinguish between the two types of flow in an experiment. In contrast, the velocity profile in a viscous flow with the no-slip boundary condition is significantly different from the Ohmic limit: it is convex in the former and concave in the latter case.

At a fixed electron density $n$, the electron transport mean free path $l_{tr}$ depends on temperature due to the electron scattering off phonons; resistivity $\rho$ is inversely proportional to $l_{tr}$. The viscosity $\eta$ of electron liquid is controlled by the electron-electron scattering and is a function of temperature as well. The competition between the viscous and Ohmic flows determines the width $l$ of the boundary layer in the electron liquid moving around an obstacle [see Eq.~(\ref{l_param})]; $l$ is proportional to $\sqrt{\eta/\rho}$ and also is a function of temperature. The crossover from Ohmic to viscous flow upon lowering the temperature occurs once $l_{tr}$ or $l$ exceeds the width $2w$ of the slit. The former case was alluded to in Ref.~[\onlinecite{Guo3068}]. Our work investigates the details of Ohmic-to-viscous crossover in the latter case (interplay between $l$ and $w$). We develop a method based on a solution of the integral equation~(\ref{integral_eq}), which depends on the parameter $l$  and describes the crossover.  We find an efficient numerical scheme to solve that equation and establish that the crossover occurs at $l/w \simeq 0.5$. In addition, by dropping certain term in the kernel $K(x)$ of the integral equation and solving it analytically, we produce a convenient extrapolation formula [see Eq.~(\ref{analytical_vx})]. The crossover is marked by the change in the current profile from concave to a convex one. 

The profile evolves slowly with the ratio $l/w$ and is rather flat at $l/w = 0.5$ (see Fig.~\ref{fig:crossover}). On the other hand, at a sufficiently low temperature, the electron transport becomes ballistic, which also leads to a flat current profile. That raises the question about the width of the temperature window in which viscous flow dominates the transport allowing the convex current profile to develop. This question is addressed in Sec.~\ref{Sec:parameterdomain}, which may help to optimize the choice of electron densities and slit widths in future experiments.

We focused on the distribution of the current density in the absence of a magnetic field. Applying it affects the spatial profiles of the electric field and current density. The magnetic-field-induced modifications to the electric potential landscape and current density around an injection point were evaluated in Ref.~[\onlinecite{Pellegrino2018}]. The results of the hydrodynamic theory in this case weakly depend on the type of the boundary condition. A channel geometry was investigated within a more microscopic approach based on the kinetic equation\cite{Holder2019}. That theory informed the experiment~\cite{ShahalIlani2019} which, in turn, indicated that the boundary condition falls in between the no-slip and no-stress limits. Theory\cite{Holder2019} also indicated that the crossover between the hydrodynamic and ballistic regimes is quite broad for the channel geometry. In addition, for the ballistic regime the kinetic approach predicted a robust spike of the Hall field in the middle of the channel, if exactly two cyclotron orbits fit into the channel's width. This beautiful observation is reminiscent of the physics of Gantmakher-Kaner effect \cite{Kaner1968}. Works [\onlinecite{Holder2019}] and ~[\onlinecite{ShahalIlani2019}] provide a strong motivation to extend the kinetic theory, with an account for the effect of magnetic field, to a slit geometry.

\begin{acknowledgments}
We thank A. Bleszynski Jayich,  M.~Goldstein, Z.~Raines, and J. Zang for useful discussions. The work is supported by NSF DMR Grant No. 2002275 (LG) and by NSF DMR Grant No. 1810544 (AY).
\end{acknowledgments}

\bibliography{biblio}

\appendix

\section{Details on numerical solution of Eq.~(\ref{integro_differential_eq}).} \label{sec:numerical_appendix}
In this Appendix, we provide the details of a numerical solution of the integral Eq.~(\ref{integral_eq}). We rely on the Chebyshev polynomials of both first $T_n$ and second $U_n$ kind, which are well suited for solving (differential or integral) equations on a finite interval.

(i) Let us treat the principal value integral appearing in Eq.~(\ref{integro_differential_eq}). We substitute the expansion (\ref{chebyshev_expansion}) in that integral and, using Eq.~(18.17.42) of Ref.~[\onlinecite{NIST_DLMF}], obtain 
\begin{align}
    \dashint_{-w}^w dx'\,\frac{v(x')}{x-x'} = -v_c \sum_{n=1}^{\infty} c_n\,\pi\, U_{2n-1}(x/w), \label{chebyshev_integration}
\end{align}
where $U_m$ are the Chebyshev polynomials of second kind.  
In addition, we express the second derivative of the Chebyshev polynomial $U_{2m-1}$ using polynomials of lesser degrees\cite{chebyshev_derivative_ref} 
\begin{align}
    &\frac{d^2U_{2m-1}(x/w)}{dx^2} =  \label{derivative}\\
    &=\left\{\begin{array}{cc}
        \frac{8}{w^2} \sum_{n=1}^{m-1} n\,(m^2-n^2)\,U_{2n-1}(x/w), & \quad m\ge 2, \\
        0, &  \quad m = 1.
    \end{array}\right. \nonumber 
\end{align}

(ii) Let us treat the last term in Eq.~(\ref{integral_eq}). The goal is to expand that term in series of $U_{2m-1}(x/w)$. We recall the definition of $K_{reg}(x)$ in Eq.~(\ref{kernel1}), and, using parity of $v(x)$ under $x \to -x$, drop odd terms in the integrand
\begin{align}
    &\int_{-w}^w dx'\, K_{reg} (x-x')\,v(x') \label{kreg_int} \\
    &= - \int_{-w}^w dx' \,v(x') \int_0^\infty \frac{dt}{2\,l}\,\frac{\sin\left(t\, (x-x')/l\right)}{(t+\sqrt{t^2+1})^2} \nonumber \\
    &= - \int_{-w}^w dx' \,v(x') \int_0^\infty \frac{dt}{2\,l}\,\frac{\sin\left(t\, x/l\right)\cos\left(t\, x'/l\right)}{(t+\sqrt{t^2+1})^2} \nonumber
\end{align}
It allows to treat the $x$ and $x'$ parts independently. We substitute the expansion~(\ref{chebyshev_expansion}) and integrate over $x'$ using the identity 
\begin{align}
\int_{-w}^w  dx' \frac{\cos(tx'/l)\,T_{2m}(x'/w)}{\sqrt{w^2-{x'}^2}}  = (-1)^m J_{2m}(t w/l). \label{aux_id1}
\end{align}
Further, we expand 
\begin{align}
  \sin(tx/l) = \frac{4\,l}{t\,w} \sum_{n=1}^\infty (-1)^{n+1}\,n \,J_{2n}(tw/l)\, U_{2n-1}(x/w). \label{aux_id2}
\end{align}
Equations~(\ref{aux_id1}) and (\ref{aux_id2}) allow to  cast Eq.~(\ref{kreg_int}) in a concise form
\begin{align}
    &\int_{-w}^w dx'\, K_{reg} (x-x')\,v(x') \label{kreg}\\ 
    &\qquad\qquad\qquad = v_c \sum_{\substack{n=1\\m=0}}^{\infty} U_{2n-1}\left(\frac{x}{w}\right)K_{reg}^{nm}\, c_m, \nonumber \\
    & \qquad K_{reg}^{nm} = (-1)^{m+n}\, 2\pi n\int_0^\infty dt \frac{J_{2m}(tw/l) \,J_{2n}(tw/l)}{t\,(t+\sqrt{t^2+1})^2}. \nonumber
\end{align}
The integrals in $K^{nm}_{reg}$ are evaluated numerically. 

(iii) Equations~(\ref{chebyshev_integration}), (\ref{derivative}) and (\ref{kreg}) allow to write Eq.~(\ref{integro_differential_eq}) in the form
\begin{widetext}
\begin{align}
   \sum_{n=1}^\infty U_{2n-1}(x/w)\left\{  K_{reg}^{n0} + \sum_{m=1}^\infty \left[ \frac{16\,\pi\,l^2}{w^2} n(m^2-n^2) \theta_{mn} - \frac{3\,\pi}{2} \delta_{nm} +K_{reg}^{nm} \right]c_m \right\} = 0, \label{sum}
\end{align}
where the notation
\begin{align}
    \theta_{mn} = \left\{\begin{array}{cc} 1, & m>n, \\ 0, & m\le n,\end{array}\right. 
\end{align}
was introduced for simplicity. For reference, the three terms in the square brackets of the latter equation correspond to the three respective terms in Eq.~(\ref{integro_differential_eq}). Using the orthogonality of the Chebyshev polynomials $U_{2n-1}(x/w)$, the system of linear equations is read-off from Eq.~(\ref{sum}) 
\begin{equation}
  \sum_{m=1}^\infty \left[ \frac{16\,\pi\,l^2}{w^2} n(m^2-n^2) \theta_{mn} - \frac{3\,\pi}{2} \delta_{nm} +K_{reg}^{nm} \right]c_m = - K_{reg}^{n0}, \quad {\rm for}\,\,n = 1,2,\ldots. \label{linear_eq1}
\end{equation}
We supplement it with the boundary condition $v(\pm w) = 0$, which, given expansion~(\ref{chebyshev_expansion}) and $c_0=1$, translates into
\begin{align}
 \sum_{n=1}^\infty c_n = -1. \label{linear_eq2}
\end{align}
Equations~(\ref{linear_eq1}) and (\ref{linear_eq2}) comprise the infinite system of linear equations for the expansion coefficients $C = (c_1,c_2,\ldots)$. We solve it numerically by truncating, i.e. by setting $c_n = 0$ for $n>N$. 
\end{widetext}

\end{document}